\documentclass[
prl, twocolumn,
superscriptaddress,
nofootinbib,
 amsmath,amssymb,
 aps,
floatfix
]{revtex4-1}

\usepackage{graphicx}
\usepackage{textcomp} 
\usepackage{gensymb} 
\usepackage{dcolumn}
\usepackage{bm}
\usepackage{amsmath, amssymb}
\usepackage{graphicx}
\usepackage{dsfont}
\usepackage{float}
\usepackage{afterpage}

\usepackage{mathtools}
\usepackage{silence}
\DeclareMathOperator{\tr}{Tr}

\newcommand{\ket}[1]{\left| #1 \right\rangle}

\newcommand{\ketbra}[2]{\left|#1\middle\rangle\!\middle\langle#2\right|}

\usepackage[normalem]{ulem}
\usepackage{color}

\newcommand\m[1]{\begin{pmatrix}#1\end{pmatrix}}

\begin{document}

\preprint{APS/123-QED}

\title{Indefinite causal order in a quantum switch}

\author{K. Goswami} 
\affiliation{Centre for Engineered Quantum Systems, School of Mathematics and Physics, University of Queensland, QLD 4072 Australia}
\author{C. Giarmatzi}
\affiliation{Centre for Engineered Quantum Systems, School of Mathematics and Physics, University of Queensland, QLD 4072 Australia}
\author{M. Kewming}
\affiliation{Centre for Engineered Quantum Systems, School of Mathematics and Physics, University of Queensland, QLD 4072 Australia}
\author{F. Costa}
\affiliation{Centre for Engineered Quantum Systems, School of Mathematics and Physics, University of Queensland, QLD 4072 Australia}
\author{C. Branciard}
\affiliation{Universit\'e Grenoble Alpes, CNRS, Grenoble INP, Institut N\'eel, 38000 Grenoble, France}
\author{J. Romero}
\email{jacq.romero@gmail.com}
\affiliation{Centre for Engineered Quantum Systems, School of Mathematics and Physics, University of Queensland, QLD 4072 Australia}
\author{A. G. White}
\affiliation{Centre for Engineered Quantum Systems, School of Mathematics and Physics, University of Queensland, QLD 4072 Australia}

\date{\today}

\begin{abstract}
\noindent Quantum mechanics allows events to happen with no definite causal order: this can be verified by measuring a causal witness, in the same way that an entanglement witness verifies entanglement.
Here we realise a photonic quantum switch, where two operations, $\hat{A}$ and $\hat{B}$, act in a quantum superposition of their two possible orders. 
The operations are on the transverse spatial mode of the photons; polarisation coherently controls their order. Our implementation ensures that the operations cannot be distinguished by spatial or temporal position---further it allows qudit encoding in the target. We confirm our quantum switch has no definite causal order by constructing a causal witness and measuring its value to be 18 standard deviations beyond the definite-order bound.
\end{abstract}
\maketitle

In daily experience, it is natural to think of events happening in a fixed causal order. Strikingly, it has been proposed that quantum physics allows for nonclassical causal structures where the order of events is \emph{indefinite}~\cite{chiribella09,oreshkov12}. It has been theoretically shown that such a possibility provides an advantage for computation~\cite{araujo14}, communication complexity~\cite{feixquantum2015,Guerin2016} and other information processing tasks~\cite{chiribella12,colnaghi11,Ebler2018}. Furthermore, investigations of indefinite causal orders suggest a promising route towards a theory that combines general relativity and quantum mechanics~\cite{hardy2007towards,zych2017bell}.

\begin{figure}[!b]
\vspace{-4mm}
 \includegraphics[width=\columnwidth]{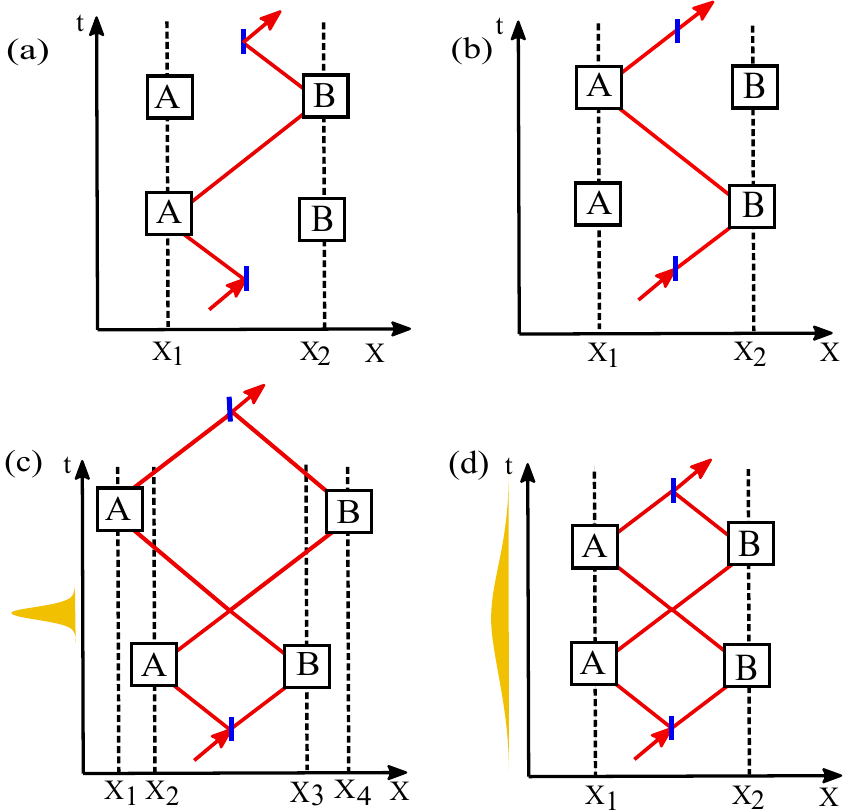}
\vspace{-5mm}
\caption{
The \emph{quantum switch}. A control qubit determines the order in which two quantum operations, $\hat{A}$ and $\hat{B}$, are applied to a target qubit, $|\psi\rangle_t$. (a) When the control is $|0\rangle_c$, $\hat{A}$ is applied  before $\hat{B}$. (b) When the control $|1\rangle_c$, $\hat{B}$ is applied before $\hat{A}$. When the control is in the superposition $(|0\rangle + |1\rangle)_c/\sqrt{2}$, there is a superposition of the two orders, yielding the output state  $|\phi\rangle {=} (\hat{B}\hat{A} |\psi\rangle_t {\otimes} |0\rangle_c {+} \hat{A}\hat{B} |\psi\rangle_t {\otimes} |1\rangle_c)/\sqrt{2}$. (c) In refs.~\cite{Procopio2014,rubino2017}, the control is the transverse position at which a photon passes through a set of wave plates---consequently, the operations are performed in distinct spatial locations depending on the order. (d) In our experiment, the control is polarisation, hence each operation takes place in a fixed spatial location, independent of the order. The yellow pulses are graphical representations of the difference in temporal characteristics: in (c) the pulses are orders-of-magnitude shorter than the experiment and its internal components; in the (d) the pulses are orders-of-magnitude longer so that the operations are indistinguishable in time as well as space.
}
\label{fig:Qswitch}
\end{figure}

Indefinite causal orders can be studied using a framework that distinguishes whether some experimental situation---called a \emph{process}---is compatible with a fixed causal order of the events or not. An example of a process with indefinite causal order is the \emph{quantum switch}~\cite{chiribella09}.  In the quantum switch, the order in which two quantum operation $\hat A$ and $\hat B$---considered as ``black box operations''---are performed on a target system is coherently controlled by a control quantum system, Fig.~\ref{fig:Qswitch}. This can also be seen as a particular case of ``superposition of time evolution" \cite{Aharonov90}. The advantages provided by the quantum switch arise from the fact that it cannot be reproduced by an ordinary quantum circuit which uses the same number of black-box operations~\cite{araujo14,feixquantum2015,chiribella12,colnaghi11,Guerin2016}.

Here we present an optical implementation of the quantum switch where the control system is the photon's polarisation and the target is the transverse spatial mode. We verify indefinite causal order by introducing a causal witness~\cite{araujo15,Branciard2016b}, for which we obtain a value 18~standard deviations beyond the bound for definite ordering.
One notable achievement of our experiment is that it opens the possibility of encoding more than two levels in the target system---transverse spatial mode can indeed be high-dimensional, and hence can act as a qudit. 

In previous implementations~\cite{Procopio2014,rubino2017}, 
the location of each black-box---the spot where photons go through a set of waveplates---was different depending on the order, resulting in four distinct locations in space%
, Fig.~\ref{fig:Qswitch}(c). 
Furthermore, the photons had a coherence length much shorter than the distance between the two sets of waveplates: in effect, the operations could also be distinct in time. 
In our experiment we use polarisation to control the order, so that the paths corresponding to different temporal orders overlap in space, resulting in only two spatial locations, as in Fig.~\ref{fig:Qswitch}(d). The location of the two operations also cannot be distinguished in time, as we use photons with a coherence length much longer than the whole interferometer. Therefore, in our experiment the causal order between operations cannot be distinguished, even in principle, by their spacetime location.

Much of the interest in the quantum switch comes from the possibility of representing a novel---genuinely quantum---type of causal structure. 
In this context, causal relations are defined through the possibility of transmitting  signals between events. An \emph{event} is understood operationally  in terms of operations such as measurements, preparations, or transformations of a physical system (for example, a photon transiting a set of lenses can define an event). A \emph{causal structure} represents the network of possible causal relations between a set of events. Relativistic causal structure naturally falls within this perspective: if an event $A$ is in the past light-cone of an event $B$ it is possible to send a signal from $A$ to $B$, while no signal exchange is allowed for space-like separated events. Note that, even if no direct causal influence between certain events is detected in a given experiment, it might still be possible to deduce information about the causal structure in which the events are embedded. For example, one can perform transformations on a system at the prescribed events and, by measuring the system at a later event, deduce information about the order in which the transformations were applied. Our experiment uses this idea.

Since the notion of causal structure refers to how different operations on physical systems relate to each other, distinct definitions are available depending on the level of description of these operations and of the physical systems. Recent literature~\cite{oreshkov12,araujo15,oreshkov15,Branciard2016b,costa2016} has analysed the quantum causal structure of the quantum switch using a theory- and device-dependent approach: we follow this here. The  possible operations defining an event $A$ are identified with the most general quantum operations: completely positive (CP) maps from an input space, $A_I {\equiv} \mathcal{L}\left(\mathcal{H}^{A_I}\right)$ to an output space, $A_O {\equiv} \mathcal{L}\left(\mathcal{H}^{A_O}\right)$, where $\mathcal{L}\left(\mathcal{H}\right)$ denotes the space of linear operators over a Hilbert space $\mathcal{H}$ and ${A_I}$, ${A_O}$ label the spaces attached to the system immediately before and after the operation, respectively. Using the Choi-Jamio{\l}kowski isomorphism~\cite{choi75,jamio72} we represent a CP map as a positive semidefinite operator $M^{A} {\in} A_I {\otimes} A_O$.
The probability to realise the maps $\left\{M^A, M^B, \dots\right\}$ in an experiment, corresponding to the events $A, B,\dots,$ is given by the \emph{generalised Born rule}~\cite{gutoski06,chiribella09b,oreshkov12,shrapnel2017} 
\begin{equation}
P(M^A,M^B,\dots)=\tr\left[\left(M^{A}\otimes M^{B} \otimes \dots\right) W\right],
\label{Born}
\end{equation}
where $W {\in} A_I {\otimes} A_O {\otimes} B_I {\otimes} B_O {\otimes} \dots$ is a positive semidefinite matrix called a \emph{process matrix}, which provides a full specification of the possible correlations that can be observed. In particular, $W$ encodes the causal structure, namely which events can potentially influence which other events.

\begin{figure}[!b]
\vspace{-7mm}
\includegraphics[width=0.8\columnwidth]{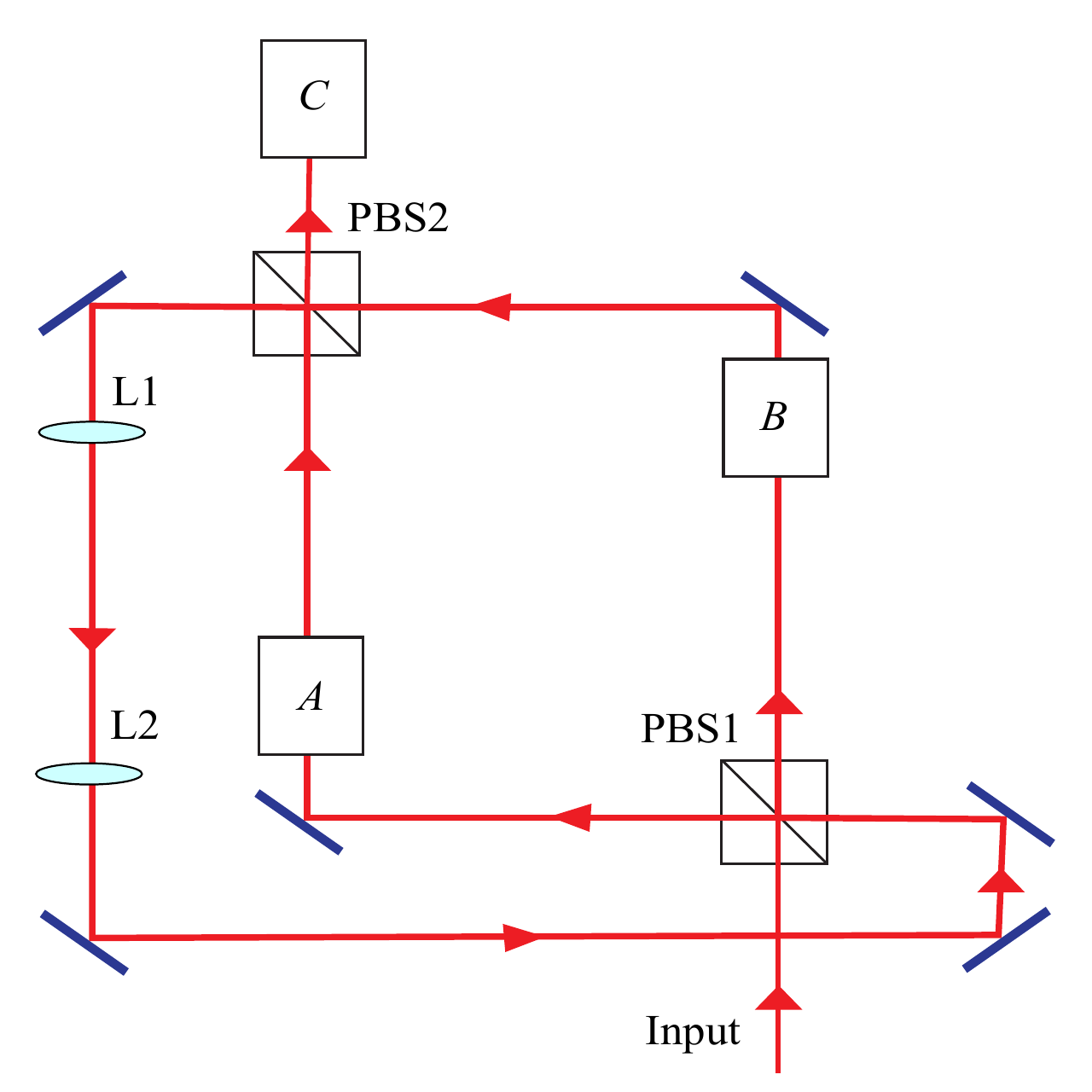}
\vspace{-3mm}
\caption{Experimental schematic. The control qubit is defined by polarisation. The polarising beamsplitter PBS1 routes the photon into either events $A$ or $B$, which realise unitary operations $\hat{A}$ or $\hat{B}$ acting on the spatial mode of the photon. Event $C$ is a $\hat{X}$ polarisation measurement, determining the Stokes parameter of the photon in the diagonal/anti-diagonal basis. Lenses L1 and L2 are used as a telescope to ensure mode-matching.
}
\label{fig:Experimental_setup1}
\end{figure}

Three events $A, B$ and $C$ must be identified in a quantum switch: $A$ and $B$ correspond to operations on the target system implemented along the two arms of the interferometer, while $C$ (not shown in Fig.~\ref{fig:Qswitch}; see Fig.~\ref{fig:Experimental_setup1}) is a measurement on the control system that occurs after both events $A$ and $B$ .
A process matrix compatible with $A$ causally preceding $B$ is denoted $W^{A\prec B\prec C}$; a process matrix with $B$ preceding $A$ is denoted $W^{B\prec A\prec C}$.
If a causal order between the events is well defined for each run of the experiment, possibly changing randomly between different runs, then the process matrix is said to be \emph{causally separable}~\cite{oreshkov12,araujo15,oreshkov15,wechs18} and it can be decomposed in the form
\vspace{-2mm}
\begin{equation}
W_{\textrm{sep}}= q \, W^{A\prec B\prec C} + (1-q) \, W^{B\prec A\prec C},
\label{separable}
\end{equation}
where $0\leq q \leq 1$.

Our task is to experimentally verify an indefinite causal order, namely that the process matrix describing our experiment cannot be decomposed as in Eq.~\eqref{separable}. We achieve this by measuring a \emph{causal witness}~\cite{araujo15,Branciard2016b}. This is defined as a Hermitian operator $S$ such that its expectation value is
\vspace{-2mm}
\begin{equation}
\langle S\rangle  = \tr [ S \, W_{\textrm{sep}} ] \geq 0
\label{witness}
\end{equation} 
for every causally separable process matrix $W_{\textrm{sep}}$. Detecting a  value $\langle S \rangle = \tr [S W] <0$ therefore certifies that $W$ is \emph{causally nonseparable}.

\begin{figure*}[!t]
\includegraphics[width=1.8\columnwidth]{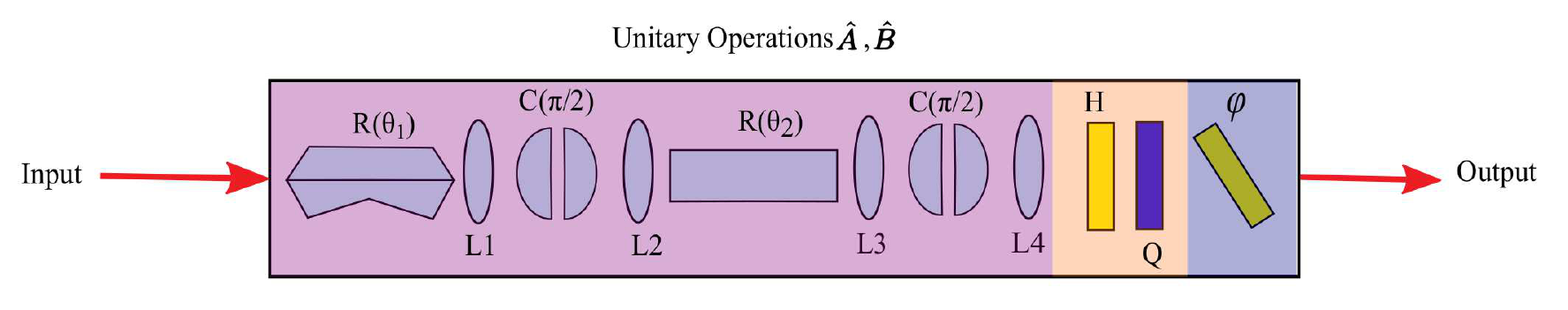}
\vspace{-3mm}
\caption{Top view of the setup for realising the unitary operations $\hat{A}$ and $\hat{B}$ using a set of special inverting prisms R, and pairs of cylindrical lenses C. The prisms rotate the incoming transverse mode, effectively implementing the rotation $\textup{R}(\theta)=\m{\rm{cos} \,2\theta & \rm{sin}\,2\theta\\\rm{sin}\,2\theta & -\rm{cos}\,2\theta}$ in the $\{ \ket{\rm{HG}_{10}}, \ket{\rm{HG}_{01}} \}$ qubit subspace. The cylindrical lenses give a $\pi/2$ relative phase shift to Hermite-Gaussian components of the incoming photon, effectively implementing $\rm{C}\,(\pi/2)=\m{1 & 0\\0 & i}$. The spherical lenses (L) are used for mode-matching. The half-waveplates (H) and quarter-waveplates (Q) are used to correct polarisation changes caused by reflections in the prisms and $\varphi$ represents a phase plate. The unitary operations of our interest are realised by varying the angles $\theta_1$ and $\theta_2$. For example, in the figure R($\theta_1$) is rotated by 45$\degree$ and for R($\theta_2$), the angle is set at 0$\degree$. With a 0$\degree$ global phase, the above setup represents an $\hat{X}$ operation which transforms an input Hermite-Gaussian HG$_{10}$ beam to a Hermite-Gaussian HG$_{01}$ (see Supplemental Material, which includes \cite{Padgett1999}). \vspace{-3mm}}
\label{fig:Experimental_setup2}
\end{figure*}

Notice that the value of $\tr [S W]$ can be obtained experimentally after decomposing $S$ into different operators representing different CP maps (which define the events $A,B,C$), and using Eq.~\eqref{Born} to write $\tr [S W]$ as a combination of the joint probabilities of these maps to be realised.
In our experiment, we use a causal witness that can be measured by letting events $A$ and $B$ correspond to unitary operations $\hat{A}$ and $\hat{B}$, respectively, and event $C$ to a polarisation measurement of the control qubit in the diagonal/antidiagonal basis---i.e., a measurement of a Stokes parameter, or equivalently of the Pauli observable $\hat{X}$. Such a witness can be decomposed as  
\begin{equation}
S = \frac14 \Big( \hat{I} + \sum_{\hat{A},\hat{B}} \gamma_{\hat{A},\hat{B}} \, \mathcal{A} \otimes \mathcal{B} \otimes \hat{X} \Big),
\label{decomposition}
\end{equation}
where $\hat{I}$ denotes the identity operator, $\mathcal{A}$ is the Choi representation of a unitary $\hat{A}$, defined as $\mathcal{A}:=( \sum_{lm} \ketbra{l}{m} \otimes \hat{A} \ketbra{l}{m}\hat{A}^{\dag} )^T$ (where $^T$ denotes transposition in the computational basis $\{\ket{l}\}$ of $\mathcal{H}^{A_I}$ \and some fixed basis of $\mathcal{H}^{A_O}$), similarly for $\mathcal{B}$, and where the sum runs over the sets of unitaries $\hat{A},\hat{B}$ used in the experiment. 
The normalisation factor $\frac14$ is chosen so that the value of $-\tr [S W]$, when positive (i.e., when $\tr [S W] < 0$), corresponds precisely to the amount of white noise that can be added to $W$ before it becomes causally separable---its ``random robustness''~\cite{araujo15,Branciard2016b}---and the coefficients $\gamma_{\hat{A},\hat{B}}$ are determined numerically through the optimisation method described in the Supplemental Material.
The value we need to measure is then
\begin{align}
\langle S\rangle &= \tr \Big[ \frac14 \Big( \hat{I} + \sum_{\hat{A},\hat{B}} \gamma_{\hat{A},\hat{B}} \, \mathcal{A} \otimes \mathcal{B} \otimes \hat{X} \Big) \, W_{\textup{exp}} \Big] \nonumber \\
 &=  1 + \frac14 \sum_{\hat{A},\hat{B}} \gamma_{\hat{A},\hat{B}} \, \langle \hat{X}\rangle_{\hat{A}, \hat{B}} \, ,
\label{probabilities}
\end{align}
where $W_{\textup{exp}}$ is the process matrix describing our experiment (properly normalised so that $\tr W_{\textup{exp}} = 4$ for our quantum switch~\cite{araujo15,oreshkov15,Branciard2016b}) and $\langle \hat{X}\rangle_{\hat{A}, \hat{B}}$ is the expectation value of the observable $\hat{X}$ given that $\hat{A}$ and $\hat{B}$ have been performed. If $\langle S\rangle$ is found to be negative, it means that our implementation of the quantum switch has successfully realised an indefinite causal order between the events $A$ and $B$.

To avoid the issue of spatial separation in Refs.~\cite{Procopio2014,rubino2017}, we use polarisation as the control qubit for the order of black boxes and the transverse spatial mode of the photons as our target qubit. With this encoding there is a single optical axis throughout the quantum switch, meaning that  each black box cannot be spatially split into two.  Furthermore, we achieve temporal indistinguishability by using photons of long coherence length. The spatial separation between the input and output of black boxes $A$ and $B$ in Fig.~\ref{fig:Experimental_setup1} is 3.5 m and two orders of magnitude shorter than the coherence length of our light source which is 955 m, such that no significant timing information can be derived. 

Our light source is a diagonally-polarised, 100 kHz linewidth laser beam at 795 nm, in the lowest-order transverse spatial mode, the Hermite-Gaussian mode HG$_{00}$. We transform the beam into a HG$_{10}$ spatial mode by first passing the beam through an element that adds a $\pi$-phase to half of the beam---a cover slip on a tip-tilt mount that spans half of the beam. The resulting spatial mode is a superposition of odd-order Hermite-Gaussian modes~\cite{Romero2012orbital}. We then use spatial Fourier filtering to remove most of  the higher-order spatial modes leaving just the HG$_{10}$ mode. The qubit space of the target system consists of first-order spatial modes, where we define $\ket{0} {=} \ket{\rm{HG}_{10}}$, and $\ket{1} {=} \ket{\rm{HG}_{01}}$. The initial state of $\ket{\psi}_t$ (Fig.~\ref{fig:Qswitch}) is taken to be $\ket{0}$. 

A polarising beamsplitter (PBS1) splits the beam into the top and bottom arms of an interferometer, see Fig.~\ref{fig:Experimental_setup1}. The unitary operations in these arms, $\hat{A}$ and $\hat{B}$, act on the transverse spatial mode, but should, ideally, not change the polarisation of the beam. The top and bottom arms are combined at the output polarising beamsplitter, PBS2, and the resulting mode is sent back to the other input of PBS1; this relay arm contains a telescope to ensure mode-matching, i.e., that the spatial mode that re-enters the interferometer is the same as the input spatial mode. 

We realise the unitary operations $\hat{A}$ and $\hat{B}$ using a combination of inverting prisms~\cite{Leach2004} and cylindrical lenses~\cite{Tamm1990,Beijersbergen1993} as shown in Fig.~\ref{fig:Experimental_setup2}. The inverting prisms rotate the incoming spatial mode.  Unlike Dove prisms which act as poor polarisers~\cite{Sullivan1972}, an inverting prism also acts approximately as a quarter-waveplate on polarisation~\cite{Leach2004}, which we compensate using a combination of quarter- and half-waveplates.

\begin{figure}[!t]
\includegraphics[width=\columnwidth]{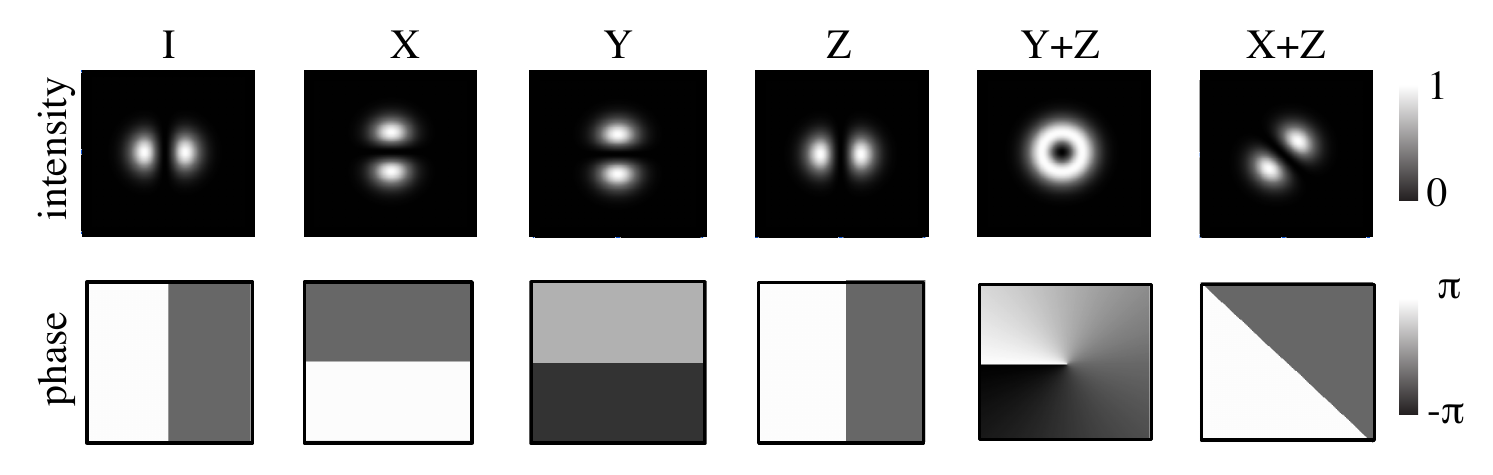}
\vspace{-6mm}
\caption{Spatial transformations. The result of the unitaries acting on an input spatial mode of HG$_{10}$ are also first-order spatial modes. \vspace{-7mm}
}
\label{fig:transformations}
\end{figure}

Transformations for spatial modes require more optical elements than transformations for polarisation,  hence in constructing the witness, we considered a tradeoff between its robustness to noise and the number of elements required to measure it in our setup. For this reason, in our experiment each operation $\hat{A}$ and $\hat{B}$ is chosen among one of the following six unitaries acting on the transverse spatial mode: the identity operation $\hat{I}$, the three Pauli operators $\hat{X}$, $\hat{Y}$  and $\hat{Z}$, and the two linear combinations $\hat{P}=(\hat{Y}+\hat{Z})/\sqrt{2}$ and $\hat{Q}=(\hat{X}+\hat{Z})/\sqrt{2}$. These operations produce spatial modes that are either first-order Hermite-Gaussian or first-order Laguerre-Gaussian modes, thus keeping the spatial mode in the $\{ \ket{\rm{HG}_{10}}, \ket{\rm{HG}_{01}} \}$ qubit subspace. Fig.~\ref{fig:transformations} illustrates the resulting spatial modes for an input target qubit in the HG$_{10}$ mode.

At the interferometer output, after PBS2, event \emph{C} corresponds to a polarisation measurement in the diagonal/antidiagonal basis---a measurement of the Stokes parameter corresponding to $\langle \hat{X}\rangle$---selected using a half-waveplate and a third polarising beamsplitter. Due to experimental imperfections in the optical elements, the output mode has a marked transverse interference pattern, typically with two to three fringes. An iris is used to collect only light from one fringe, and this is then collected by a multimode fibre connected to a single-photon detector, thus tracing out the spatial mode of the photons. 

For our witness, there are 21 combinations of $\hat{A}$ and $\hat{B}$ for which the coefficient $\gamma_{\hat{A},\hat{B}}$ is nonzero (see Supplemental Material). Fig.~\ref{fig:Stokes} shows the measured Stokes values $\langle\hat{X}\rangle_{\hat{A}, \hat{B}}$ for each of these combinations: the red bars are the theoretically expected values, which should all be $+1$, $-1$ or $0$; the blue bars are the values measured in our experiment.  

There are two main sources of errors in our experiment: rotational misalignments and imperfect mode matching. The inverting prisms are mounted on manual rotation stages with an uncertainty in angular position of $1^\circ$. Our witness is robust against these misalignments: accounting for these errors, one can derive a new corrected bound for causally separable processes, which we find to be close enough to zero that we still have room to obtain an experimental value below it (see Supplemental Material). The imperfect mode-matching degrades the visibility of the interference of the spatial modes, which is then reflected in the values of the Stokes parameters that we obtain.
We have modelled these imperfections and predict an expectation value for our causal witness within the range $-0.20 \lesssim \langle S\rangle \lesssim -0.14$, c.f. the ideal value of $\langle S\rangle \simeq -0.248$. 

\begin{figure}[!t]
\includegraphics[width=\columnwidth]{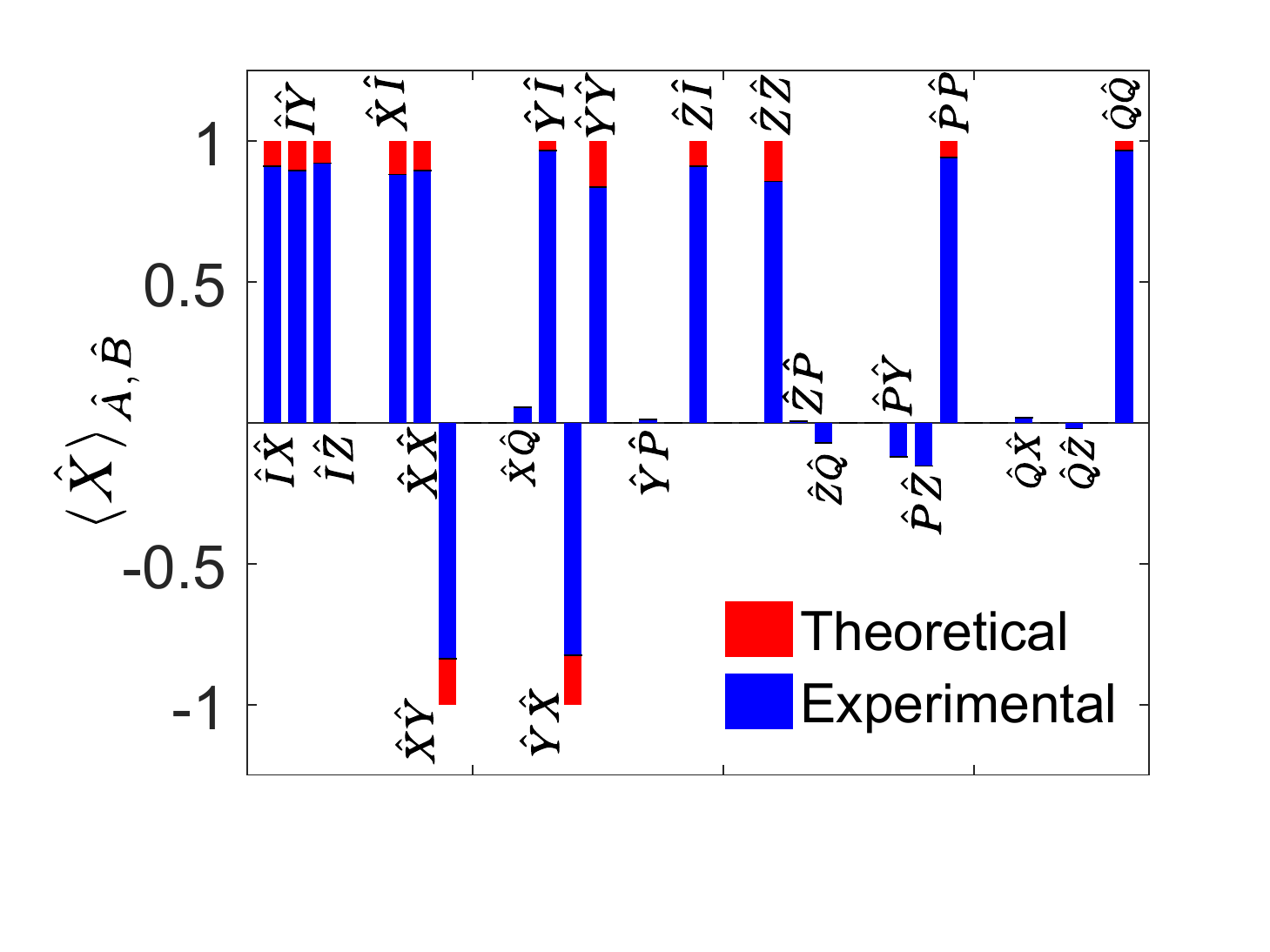}
\vspace{-17mm}
\caption{Stokes parameters $\langle \hat{X}\rangle_{\hat{A}, \hat{B}}$ obtained by measuring the polarisation of the output control qubit in the diagonal basis. The red bars show the ideal, theoretical values and the blue bars are the experimentally measured values. The unitary combinations are defined by combining the unitary operations at the top arm ($\hat{A}$) and the bottom ($\hat{B}$) arm, maintaining the order $\hat{I}$, $\hat{X}$, $\hat{Y}$, $\hat{Z}$, $\hat{P}=\frac{\hat{Y}+\hat{Z}}{\sqrt{2}}$ and $\hat{Q}=\frac{\hat{X}+\hat{Z}}{\sqrt{2}}$. A Stokes parameter of $+1$ means the output is diagonally polarised light, $-1$ means it is anti-diagonally polarised light.  1$\sigma$ errors are too small to be visible in the plot. \vspace{-5mm}}
\label{fig:Stokes}
\end{figure}

We measure $\langle S\rangle {=} {-}0.171 {\pm} 0.009$, within our expected range, and a value that is 18 standard deviations from the bound $\langle S\rangle {\ge} 0$ satisfied by all causally separable processes.  Taking into account misalignment errors, the measured value is still 14 standard deviations below the---most conservative---corrected bound of $\langle S\rangle {\ge} -0.038$ for causally separable processes (obtained in the Supplemental Material). This confirms that the measured process is causally nonseparable: it has no definite causal order.

The control and target systems in our experiment are encoded---as in previous experiments~\cite{Procopio2014,rubino2017,rubino2017b}---on different degrees of freedom of a single particle. As in all experiments, there are non-ideal aspects, which we detail in the Supplemental Material. Our experiment---and those of Refs.~\cite{Procopio2014,rubino2017}---do not overtly suffer from these non-ideal aspects as can be seen by the high visibility observed in all implementations. The high visibility ensures the target operations are sufficiently similar for different control states, and that there is no net operation on the control. 

Our architecture offers promising routes for further experimental investigations. Having polarisation as the control degree of freedom enables for instance using polarisation-entanglement---which can be of very high quality, e.g. reaching a tangle of $T \simeq 0.987$~\cite{Fedrizzi:07}---as the control for entangling the causal order of different quantum switches~\cite{zych2017bell,rubino2017b}. Having transverse spatial modes as the target degree of freedom enables encoding qudits---as opposed to qubits---for investigating quantum communication with indefinite causal order in larger Hilbert space dimensions~\cite{Guerin2016,Ebler2018}. The benefits of using qudits for quantum information processing applications, such as improved security in quantum cryptography and higher information capacity in quantum communication, are well-known \cite{erhard2018twisted,cerf2002security}.  Moreover, certain protocols demonstrating an advantage from indefinite causal order will require qudits for their implementation~\cite{Guerin2016,Ebler2018}. Our implementation thus offers the possibility of exploring these advantages in the future. Other challenges  include realising quantum switches that put more than two events in an indefinite causal order, and physically separating the control and target systems, so that the parties' actions on the target system for different control states cannot be distinguished even in principle.  

\noindent \emph{Acknowledgements}. We thank A. A. Abbott for useful feedback on our manuscript. This work has been supported by: the Australian Research Council (ARC) by DECRA grants for FC (DE170100712) and JR (DE160100409), and the Centre of Excellence for Engineered Quantum Systems (EQUS, CE170100009); L'Oreal-UNESCO For Women In Science grant for JR; the French National Research Agency by the ``Retour Post-Doctorants'' program (ANR-13-PDOC-0026) for CB; the University of Queensland by a Vice-Chancellors Senior Research and Teaching Fellowship for AGW; and the John Templeton Foundation (the opinions expressed in this publication are those of the authors and do not necessarily reflect the views of the John Templeton Foundation). We acknowledge the traditional owners of the land on which the University of Queensland is situated, the Turrbal and Jagera people.  
\vspace{-5mm}

\begin{thebibliography}{33}%
\makeatletter
\providecommand \@ifxundefined [1]{%
 \@ifx{#1\undefined}
}%
\providecommand \@ifnum [1]{%
 \ifnum #1\expandafter \@firstoftwo
 \else \expandafter \@secondoftwo
 \fi
}%
\providecommand \@ifx [1]{%
 \ifx #1\expandafter \@firstoftwo
 \else \expandafter \@secondoftwo
 \fi
}%
\providecommand \natexlab [1]{#1}%
\providecommand \enquote  [1]{``#1''}%
\providecommand \bibnamefont  [1]{#1}%
\providecommand \bibfnamefont [1]{#1}%
\providecommand \citenamefont [1]{#1}%
\providecommand \href@noop [0]{\@secondoftwo}%
\providecommand \href [0]{\begingroup \@sanitize@url \@href}%
\providecommand \@href[1]{\@@startlink{#1}\@@href}%
\providecommand \@@href[1]{\endgroup#1\@@endlink}%
\providecommand \@sanitize@url [0]{\catcode `\\12\catcode `\$12\catcode
  `\&12\catcode `\#12\catcode `\^12\catcode `\_12\catcode `\%12\relax}%
\providecommand \@@startlink[1]{}%
\providecommand \@@endlink[0]{}%
\providecommand \url  [0]{\begingroup\@sanitize@url \@url }%
\providecommand \@url [1]{\endgroup\@href {#1}{\urlprefix }}%
\providecommand \urlprefix  [0]{URL }%
\providecommand \Eprint [0]{\href }%
\providecommand \doibase [0]{http://dx.doi.org/}%
\providecommand \selectlanguage [0]{\@gobble}%
\providecommand \bibinfo  [0]{\@secondoftwo}%
\providecommand \bibfield  [0]{\@secondoftwo}%
\providecommand \translation [1]{[#1]}%
\providecommand \BibitemOpen [0]{}%
\providecommand \bibitemStop [0]{}%
\providecommand \bibitemNoStop [0]{.\EOS\space}%
\providecommand \EOS [0]{\spacefactor3000\relax}%
\providecommand \BibitemShut  [1]{\csname bibitem#1\endcsname}%
\let\auto@bib@innerbib\@empty
\bibitem [{\citenamefont {{Chiribella}}\ \emph {et~al.}(2013)\citenamefont
  {{Chiribella}}, \citenamefont {{D'Ariano}}, \citenamefont {{Perinotti}},\
  and\ \citenamefont {{Valiron}}}]{chiribella09}%
  \BibitemOpen
  \bibfield  {author} {\bibinfo {author} {\bibfnamefont {G.}~\bibnamefont
  {{Chiribella}}}, \bibinfo {author} {\bibfnamefont {G.~M.}\ \bibnamefont
  {{D'Ariano}}}, \bibinfo {author} {\bibfnamefont {P.}~\bibnamefont
  {{Perinotti}}}, \ and\ \bibinfo {author} {\bibfnamefont {B.}~\bibnamefont
  {{Valiron}}},\ }\href {\doibase 10.1103/PhysRevA.88.022318} {\bibfield
  {journal} {\bibinfo  {journal} {Phys. Rev.~A}\ }\textbf {\bibinfo {volume}
  {88}},\ \bibinfo {eid} {022318} (\bibinfo {year} {2013})},\ \Eprint
  {http://arxiv.org/abs/0912.0195} {arXiv:0912.0195 [quant-ph]} \BibitemShut
  {NoStop}%
\bibitem [{\citenamefont {{Oreshkov}}\ \emph {et~al.}(2012)\citenamefont
  {{Oreshkov}}, \citenamefont {{Costa}},\ and\ \citenamefont
  {{Brukner}}}]{oreshkov12}%
  \BibitemOpen
  \bibfield  {author} {\bibinfo {author} {\bibfnamefont {O.}~\bibnamefont
  {{Oreshkov}}}, \bibinfo {author} {\bibfnamefont {F.}~\bibnamefont {{Costa}}},
  \ and\ \bibinfo {author} {\bibfnamefont {{\v C}.}~\bibnamefont {{Brukner}}},\
  }\href {\doibase 10.1038/ncomms2076} {\bibfield  {journal} {\bibinfo
  {journal} {Nat. Commun.}\ }\textbf {\bibinfo {volume} {3}},\ \bibinfo {pages}
  {1092} (\bibinfo {year} {2012})},\ \Eprint {http://arxiv.org/abs/1105.4464}
  {arXiv:1105.4464 [quant-ph]} \BibitemShut {NoStop}%
\bibitem [{\citenamefont {{Ara{\'u}jo}}\ \emph {et~al.}(2014)\citenamefont
  {{Ara{\'u}jo}}, \citenamefont {{Costa}},\ and\ \citenamefont
  {{Brukner}}}]{araujo14}%
  \BibitemOpen
  \bibfield  {author} {\bibinfo {author} {\bibfnamefont {M.}~\bibnamefont
  {{Ara{\'u}jo}}}, \bibinfo {author} {\bibfnamefont {F.}~\bibnamefont
  {{Costa}}}, \ and\ \bibinfo {author} {\bibfnamefont {{\v C}.}~\bibnamefont
  {{Brukner}}},\ }\href {\doibase 10.1103/PhysRevLett.113.250402} {\bibfield
  {journal} {\bibinfo  {journal} {Phys. Rev. Lett.}\ }\textbf {\bibinfo
  {volume} {113}},\ \bibinfo {pages} {250402} (\bibinfo {year} {2014})},\
  \Eprint {http://arxiv.org/abs/1401.8127} {arXiv:1401.8127 [quant-ph]}
  \BibitemShut {NoStop}%
\bibitem [{\citenamefont {Feix}\ \emph {et~al.}(2015)\citenamefont {Feix},
  \citenamefont {Ara\'ujo},\ and\ \citenamefont {Brukner}}]{feixquantum2015}%
  \BibitemOpen
  \bibfield  {author} {\bibinfo {author} {\bibfnamefont {A.}~\bibnamefont
  {Feix}}, \bibinfo {author} {\bibfnamefont {M.}~\bibnamefont {Ara\'ujo}}, \
  and\ \bibinfo {author} {\bibfnamefont {{\v C}.}~\bibnamefont {Brukner}},\
  }\href {\doibase 10.1103/PhysRevA.92.052326} {\bibfield  {journal} {\bibinfo
  {journal} {Phys. Rev. A}\ }\textbf {\bibinfo {volume} {92}},\ \bibinfo
  {pages} {052326} (\bibinfo {year} {2015})},\ \Eprint
  {http://arxiv.org/abs/1508.07840} {arXiv:1508.07840 [quant-ph]} \BibitemShut
  {NoStop}%
\bibitem [{\citenamefont {Gu{\'e}rin}\ \emph {et~al.}(2016)\citenamefont
  {Gu{\'e}rin}, \citenamefont {Feix}, \citenamefont {Ara{\'u}jo},\ and\
  \citenamefont {Brukner}}]{Guerin2016}%
  \BibitemOpen
  \bibfield  {author} {\bibinfo {author} {\bibfnamefont {P.~A.}\ \bibnamefont
  {Gu{\'e}rin}}, \bibinfo {author} {\bibfnamefont {A.}~\bibnamefont {Feix}},
  \bibinfo {author} {\bibfnamefont {M.}~\bibnamefont {Ara{\'u}jo}}, \ and\
  \bibinfo {author} {\bibfnamefont {{\v C}.}~\bibnamefont {Brukner}},\ }\href
  {\doibase 10.1103/PhysRevLett.117.100502} {\bibfield  {journal} {\bibinfo
  {journal} {Phys. Rev. Lett.}\ }\textbf {\bibinfo {volume} {117}},\ \bibinfo
  {pages} {100502} (\bibinfo {year} {2016})},\ \Eprint
  {http://arxiv.org/abs/1605.07372} {arXiv:1605.07372 [quant-ph]} \BibitemShut
  {NoStop}%
\bibitem [{\citenamefont {{Chiribella}}(2012)}]{chiribella12}%
  \BibitemOpen
  \bibfield  {author} {\bibinfo {author} {\bibfnamefont {G.}~\bibnamefont
  {{Chiribella}}},\ }\href {\doibase 10.1103/PhysRevA.86.040301} {\bibfield
  {journal} {\bibinfo  {journal} {Phys. Rev.~A}\ }\textbf {\bibinfo {volume}
  {86}},\ \bibinfo {eid} {040301} (\bibinfo {year} {2012})},\ \Eprint
  {http://arxiv.org/abs/1109.5154} {arXiv:1109.5154 [quant-ph]} \BibitemShut
  {NoStop}%
\bibitem [{\citenamefont {{Colnaghi}}\ \emph {et~al.}(2012)\citenamefont
  {{Colnaghi}}, \citenamefont {{D'Ariano}}, \citenamefont {{Facchini}},\ and\
  \citenamefont {{Perinotti}}}]{colnaghi11}%
  \BibitemOpen
  \bibfield  {author} {\bibinfo {author} {\bibfnamefont {T.}~\bibnamefont
  {{Colnaghi}}}, \bibinfo {author} {\bibfnamefont {G.~M.}\ \bibnamefont
  {{D'Ariano}}}, \bibinfo {author} {\bibfnamefont {S.}~\bibnamefont
  {{Facchini}}}, \ and\ \bibinfo {author} {\bibfnamefont {P.}~\bibnamefont
  {{Perinotti}}},\ }\href {\doibase 10.1016/j.physleta.2012.08.028} {\bibfield
  {journal} {\bibinfo  {journal} {Phys. Lett.~A}\ }\textbf {\bibinfo {volume}
  {376}},\ \bibinfo {pages} {2940} (\bibinfo {year} {2012})},\ \Eprint
  {http://arxiv.org/abs/1109.5987} {arXiv:1109.5987 [quant-ph]} \BibitemShut
  {NoStop}%
\bibitem [{\citenamefont {Ebler}\ \emph {et~al.}(2018)\citenamefont {Ebler},
  \citenamefont {Salek},\ and\ \citenamefont {Chiribella}}]{Ebler2018}%
  \BibitemOpen
  \bibfield  {author} {\bibinfo {author} {\bibfnamefont {D.}~\bibnamefont
  {Ebler}}, \bibinfo {author} {\bibfnamefont {S.}~\bibnamefont {Salek}}, \ and\
  \bibinfo {author} {\bibfnamefont {G.}~\bibnamefont {Chiribella}},\ }\href
  {\doibase 10.1103/PhysRevLett.120.120502} {\bibfield  {journal} {\bibinfo
  {journal} {Phys. Rev. Lett.}\ }\textbf {\bibinfo {volume} {120}},\ \bibinfo
  {pages} {120502} (\bibinfo {year} {2018})},\ \Eprint
  {http://arxiv.org/abs/1711.10165} {arXiv:1711.10165 [quant-ph]} \BibitemShut
  {NoStop}%
\bibitem [{\citenamefont {{Hardy}}(2007)}]{hardy2007towards}%
  \BibitemOpen
  \bibfield  {author} {\bibinfo {author} {\bibfnamefont {L.}~\bibnamefont
  {{Hardy}}},\ }\href {\doibase 10.1088/1751-8113/40/12/S12} {\bibfield
  {journal} {\bibinfo  {journal} {J.~Phys. A: Math. Gen.}\ }\textbf {\bibinfo
  {volume} {40}},\ \bibinfo {pages} {3081} (\bibinfo {year} {2007})},\ \Eprint
  {http://arxiv.org/abs/gr-qc/0608043} {arXiv:gr-qc/0608043} \BibitemShut
  {NoStop}%
\bibitem [{\citenamefont {Zych}\ \emph {et~al.}(2017)\citenamefont {Zych},
  \citenamefont {Costa}, \citenamefont {Pikovski},\ and\ \citenamefont
  {Brukner}}]{zych2017bell}%
  \BibitemOpen
  \bibfield  {author} {\bibinfo {author} {\bibfnamefont {M.}~\bibnamefont
  {Zych}}, \bibinfo {author} {\bibfnamefont {F.}~\bibnamefont {Costa}},
  \bibinfo {author} {\bibfnamefont {I.}~\bibnamefont {Pikovski}}, \ and\
  \bibinfo {author} {\bibfnamefont {{\v C}.}~\bibnamefont {Brukner}},\
  }\href@noop {} {\  (\bibinfo {year} {2017})},\ \Eprint
  {http://arxiv.org/abs/1708.00248} {arXiv:1708.00248 [quant-ph]} \BibitemShut
  {NoStop}%
\bibitem [{\citenamefont {Procopio}\ \emph {et~al.}(2015)\citenamefont
  {Procopio}, \citenamefont {Moqanaki}, \citenamefont {Ara{\'u}jo},
  \citenamefont {Costa}, \citenamefont {Calafell}, \citenamefont {Dowd},
  \citenamefont {Hamel}, \citenamefont {Rozema}, \citenamefont {Brukner},\ and\
  \citenamefont {Walther}}]{Procopio2014}%
  \BibitemOpen
  \bibfield  {author} {\bibinfo {author} {\bibfnamefont {L.~M.}\ \bibnamefont
  {Procopio}}, \bibinfo {author} {\bibfnamefont {A.}~\bibnamefont {Moqanaki}},
  \bibinfo {author} {\bibfnamefont {M.}~\bibnamefont {Ara{\'u}jo}}, \bibinfo
  {author} {\bibfnamefont {F.}~\bibnamefont {Costa}}, \bibinfo {author}
  {\bibfnamefont {I.~A.}\ \bibnamefont {Calafell}}, \bibinfo {author}
  {\bibfnamefont {E.~G.}\ \bibnamefont {Dowd}}, \bibinfo {author}
  {\bibfnamefont {D.~R.}\ \bibnamefont {Hamel}}, \bibinfo {author}
  {\bibfnamefont {L.~A.}\ \bibnamefont {Rozema}}, \bibinfo {author}
  {\bibfnamefont {{\v C}.}~\bibnamefont {Brukner}}, \ and\ \bibinfo {author}
  {\bibfnamefont {P.}~\bibnamefont {Walther}},\ }\href {\doibase
  10.1038/ncomms8913} {\bibfield  {journal} {\bibinfo  {journal} {Nat.
  Commun.}\ }\textbf {\bibinfo {volume} {6}},\ \bibinfo {pages} {7913}
  (\bibinfo {year} {2015})},\ \Eprint {http://arxiv.org/abs/1412.4006}
  {arXiv:1412.4006 [quant-ph]} \BibitemShut {NoStop}%
\bibitem [{\citenamefont {Rubino}\ \emph
  {et~al.}(2017{\natexlab{a}})\citenamefont {Rubino}, \citenamefont {Rozema},
  \citenamefont {Feix}, \citenamefont {Ara{\'u}jo}, \citenamefont {Zeuner},
  \citenamefont {Procopio}, \citenamefont {Brukner},\ and\ \citenamefont
  {Walther}}]{rubino2017}%
  \BibitemOpen
  \bibfield  {author} {\bibinfo {author} {\bibfnamefont {G.}~\bibnamefont
  {Rubino}}, \bibinfo {author} {\bibfnamefont {L.~A.}\ \bibnamefont {Rozema}},
  \bibinfo {author} {\bibfnamefont {A.}~\bibnamefont {Feix}}, \bibinfo {author}
  {\bibfnamefont {M.}~\bibnamefont {Ara{\'u}jo}}, \bibinfo {author}
  {\bibfnamefont {J.~M.}\ \bibnamefont {Zeuner}}, \bibinfo {author}
  {\bibfnamefont {L.~M.}\ \bibnamefont {Procopio}}, \bibinfo {author}
  {\bibfnamefont {{\v{C}}.}~\bibnamefont {Brukner}}, \ and\ \bibinfo {author}
  {\bibfnamefont {P.}~\bibnamefont {Walther}},\ }\href@noop {} {\bibfield
  {journal} {\bibinfo  {journal} {Sci. Adv.}\ }\textbf {\bibinfo {volume}
  {3}},\ \bibinfo {pages} {e1602589} (\bibinfo {year} {2017}{\natexlab{a}})},\
  \Eprint {http://arxiv.org/abs/1608.01683} {arXiv:1608.01683 [quant-ph]}
  \BibitemShut {NoStop}%
\bibitem [{\citenamefont {{Aharonov}}\ \emph {et~al.}(1990)\citenamefont
  {{Aharonov}}, \citenamefont {{Anandan}}, \citenamefont {{Popescu}},\ and\
  \citenamefont {{Vaidman}}}]{Aharonov90}%
  \BibitemOpen
  \bibfield  {author} {\bibinfo {author} {\bibfnamefont {Y.}~\bibnamefont
  {{Aharonov}}}, \bibinfo {author} {\bibfnamefont {J.}~\bibnamefont
  {{Anandan}}}, \bibinfo {author} {\bibfnamefont {S.}~\bibnamefont
  {{Popescu}}}, \ and\ \bibinfo {author} {\bibfnamefont {L.}~\bibnamefont
  {{Vaidman}}},\ }\href {\doibase 10.1103/PhysRevLett.64.2965} {\bibfield
  {journal} {\bibinfo  {journal} {Phys. Rev. Lett.}\ }\textbf {\bibinfo
  {volume} {64}},\ \bibinfo {pages} {2965} (\bibinfo {year}
  {1990})}\BibitemShut {NoStop}%
\bibitem [{\citenamefont {Ara{\'u}jo}\ \emph {et~al.}(2015)\citenamefont
  {Ara{\'u}jo}, \citenamefont {Branciard}, \citenamefont {Costa}, \citenamefont
  {Feix}, \citenamefont {Giarmatzi},\ and\ \citenamefont {Brukner}}]{araujo15}%
  \BibitemOpen
  \bibfield  {author} {\bibinfo {author} {\bibfnamefont {M.}~\bibnamefont
  {Ara{\'u}jo}}, \bibinfo {author} {\bibfnamefont {C.}~\bibnamefont
  {Branciard}}, \bibinfo {author} {\bibfnamefont {F.}~\bibnamefont {Costa}},
  \bibinfo {author} {\bibfnamefont {A.}~\bibnamefont {Feix}}, \bibinfo {author}
  {\bibfnamefont {C.}~\bibnamefont {Giarmatzi}}, \ and\ \bibinfo {author}
  {\bibfnamefont {{\v C}.}~\bibnamefont {Brukner}},\ }\href {\doibase
  10.1088/1367-2630/17/10/102001} {\bibfield  {journal} {\bibinfo  {journal}
  {New J. Phys.}\ }\textbf {\bibinfo {volume} {17}},\ \bibinfo {pages} {102001}
  (\bibinfo {year} {2015})},\ \Eprint {http://arxiv.org/abs/1506.03776}
  {arXiv:1506.03776 [quant-ph]} \BibitemShut {NoStop}%
\bibitem [{\citenamefont {Branciard}(2016)}]{Branciard2016b}%
  \BibitemOpen
  \bibfield  {author} {\bibinfo {author} {\bibfnamefont {C.}~\bibnamefont
  {Branciard}},\ }\href@noop {} {\bibfield  {journal} {\bibinfo  {journal}
  {Sci. Rep.}\ }\textbf {\bibinfo {volume} {6}},\ \bibinfo {pages} {26018}
  (\bibinfo {year} {2016})},\ \Eprint {http://arxiv.org/abs/1603.00043}
  {arXiv:1603.00043 [quant-ph]} \BibitemShut {NoStop}%
\bibitem [{\citenamefont {Oreshkov}\ and\ \citenamefont
  {Giarmatzi}(2016)}]{oreshkov15}%
  \BibitemOpen
  \bibfield  {author} {\bibinfo {author} {\bibfnamefont {O.}~\bibnamefont
  {Oreshkov}}\ and\ \bibinfo {author} {\bibfnamefont {C.}~\bibnamefont
  {Giarmatzi}},\ }\href {\doibase 10.1088/1367-2630/18/9/093020} {\bibfield
  {journal} {\bibinfo  {journal} {New J. Phys.}\ }\textbf {\bibinfo {volume}
  {18}},\ \bibinfo {pages} {093020} (\bibinfo {year} {2016})},\ \Eprint
  {http://arxiv.org/abs/1506.05449} {arXiv:1506.05449 [quant-ph]} \BibitemShut
  {NoStop}%
\bibitem [{\citenamefont {Costa}\ and\ \citenamefont
  {Shrapnel}(2016)}]{costa2016}%
  \BibitemOpen
  \bibfield  {author} {\bibinfo {author} {\bibfnamefont {F.}~\bibnamefont
  {Costa}}\ and\ \bibinfo {author} {\bibfnamefont {S.}~\bibnamefont
  {Shrapnel}},\ }\href {\doibase https://doi.org/10.1088/1367-2630/18/6/063032}
  {\bibfield  {journal} {\bibinfo  {journal} {New J. Phys.}\ }\textbf {\bibinfo
  {volume} {18}},\ \bibinfo {pages} {063032} (\bibinfo {year} {2016})},\
  \Eprint {http://arxiv.org/abs/1512.07106} {arXiv:1512.07106 [quant-ph]}
  \BibitemShut {NoStop}%
\bibitem [{\citenamefont {Choi}(1975)}]{choi75}%
  \BibitemOpen
  \bibfield  {author} {\bibinfo {author} {\bibfnamefont {M.-D.}\ \bibnamefont
  {Choi}},\ }\href {\doibase 10.1016/0024-3795(75)90058-0} {\bibfield
  {journal} {\bibinfo  {journal} {Linear Algebra Appl.}\ }\textbf {\bibinfo
  {volume} {12}},\ \bibinfo {pages} {95} (\bibinfo {year} {1975})}\BibitemShut
  {NoStop}%
\bibitem [{\citenamefont {Jamio{\l}kowski}(1972)}]{jamio72}%
  \BibitemOpen
  \bibfield  {author} {\bibinfo {author} {\bibfnamefont {A.}~\bibnamefont
  {Jamio{\l}kowski}},\ }\href {\doibase 10.1016/0034-4877(72)90011-0}
  {\bibfield  {journal} {\bibinfo  {journal} {Rep. Math. Phys.}\ }\textbf
  {\bibinfo {volume} {3}},\ \bibinfo {pages} {275} (\bibinfo {year}
  {1972})}\BibitemShut {NoStop}%
\bibitem [{\citenamefont {Gutoski}\ and\ \citenamefont
  {Watrous}(2006)}]{gutoski06}%
  \BibitemOpen
  \bibfield  {author} {\bibinfo {author} {\bibfnamefont {G.}~\bibnamefont
  {Gutoski}}\ and\ \bibinfo {author} {\bibfnamefont {J.}~\bibnamefont
  {Watrous}},\ }in\ \href@noop {} {\emph {\bibinfo {booktitle} {Proceedings of
  39th ACM STOC}}}\ (\bibinfo {year} {2006})\ pp.\ \bibinfo {pages}
  {565--574},\ \Eprint {http://arxiv.org/abs/quant-ph/0611234}
  {arXiv:quant-ph/0611234} \BibitemShut {NoStop}%
\bibitem [{\citenamefont {{Chiribella}}\ \emph {et~al.}(2009)\citenamefont
  {{Chiribella}}, \citenamefont {{D'Ariano}},\ and\ \citenamefont
  {{Perinotti}}}]{chiribella09b}%
  \BibitemOpen
  \bibfield  {author} {\bibinfo {author} {\bibfnamefont {G.}~\bibnamefont
  {{Chiribella}}}, \bibinfo {author} {\bibfnamefont {G.~M.}\ \bibnamefont
  {{D'Ariano}}}, \ and\ \bibinfo {author} {\bibfnamefont {P.}~\bibnamefont
  {{Perinotti}}},\ }\href {\doibase 10.1103/PhysRevA.80.022339} {\bibfield
  {journal} {\bibinfo  {journal} {Phys. Rev.~A}\ }\textbf {\bibinfo {volume}
  {80}},\ \bibinfo {eid} {022339} (\bibinfo {year} {2009})},\ \Eprint
  {http://arxiv.org/abs/0904.4483} {arXiv:0904.4483 [quant-ph]} \BibitemShut
  {NoStop}%
\bibitem [{\citenamefont {Shrapnel}\ \emph {et~al.}(2017)\citenamefont
  {Shrapnel}, \citenamefont {Costa},\ and\ \citenamefont
  {Milburn}}]{shrapnel2017}%
  \BibitemOpen
  \bibfield  {author} {\bibinfo {author} {\bibfnamefont {S.}~\bibnamefont
  {Shrapnel}}, \bibinfo {author} {\bibfnamefont {F.}~\bibnamefont {Costa}}, \
  and\ \bibinfo {author} {\bibfnamefont {G.}~\bibnamefont {Milburn}},\
  }\href@noop {} {\  (\bibinfo {year} {2017})},\ \Eprint
  {http://arxiv.org/abs/1702.01845} {arXiv:1702.01845 [quant-ph]} \BibitemShut
  {NoStop}%
\bibitem [{\citenamefont {{Wechs}}\ \emph {et~al.}(2018)\citenamefont
  {{Wechs}}, \citenamefont {{Abbott}},\ and\ \citenamefont
  {{Branciard}}}]{wechs18}%
  \BibitemOpen
  \bibfield  {author} {\bibinfo {author} {\bibfnamefont {J.}~\bibnamefont
  {{Wechs}}}, \bibinfo {author} {\bibfnamefont {A.~A.}\ \bibnamefont
  {{Abbott}}}, \ and\ \bibinfo {author} {\bibfnamefont {C.}~\bibnamefont
  {{Branciard}}},\ }\href@noop {} {\  (\bibinfo {year} {2018})},\ \Eprint
  {http://arxiv.org/abs/1807.10557} {arXiv:1807.10557 [quant-ph]} \BibitemShut
  {NoStop}%
\bibitem [{\citenamefont {Padgett}\ and\ \citenamefont
  {Lesso}(1999)}]{Padgett1999}%
  \BibitemOpen
  \bibfield  {author} {\bibinfo {author} {\bibfnamefont {M.~J.}\ \bibnamefont
  {Padgett}}\ and\ \bibinfo {author} {\bibfnamefont {J.~P.}\ \bibnamefont
  {Lesso}},\ }\href@noop {} {\bibfield  {journal} {\bibinfo  {journal} {J. Mod.
  Opt.}\ }\textbf {\bibinfo {volume} {46}},\ \bibinfo {pages} {175} (\bibinfo
  {year} {1999})}\BibitemShut {NoStop}%
\bibitem [{\citenamefont {Romero}\ \emph {et~al.}(2012)\citenamefont {Romero},
  \citenamefont {Giovannini}, \citenamefont {McLaren}, \citenamefont {Galvez},
  \citenamefont {Forbes},\ and\ \citenamefont {Padgett}}]{Romero2012orbital}%
  \BibitemOpen
  \bibfield  {author} {\bibinfo {author} {\bibfnamefont {J.}~\bibnamefont
  {Romero}}, \bibinfo {author} {\bibfnamefont {D.}~\bibnamefont {Giovannini}},
  \bibinfo {author} {\bibfnamefont {M.~G.}\ \bibnamefont {McLaren}}, \bibinfo
  {author} {\bibfnamefont {E.~J.}\ \bibnamefont {Galvez}}, \bibinfo {author}
  {\bibfnamefont {A.}~\bibnamefont {Forbes}}, \ and\ \bibinfo {author}
  {\bibfnamefont {M.~J.}\ \bibnamefont {Padgett}},\ }\href@noop {} {\bibfield
  {journal} {\bibinfo  {journal} {J. Opt.}\ }\textbf {\bibinfo {volume} {14}},\
  \bibinfo {pages} {085401} (\bibinfo {year} {2012})}\BibitemShut {NoStop}%
\bibitem [{\citenamefont {Leach}\ \emph {et~al.}(2004)\citenamefont {Leach},
  \citenamefont {Courtial}, \citenamefont {Skeldon}, \citenamefont {Barnett},
  \citenamefont {Franke-Arnold},\ and\ \citenamefont {Padgett}}]{Leach2004}%
  \BibitemOpen
  \bibfield  {author} {\bibinfo {author} {\bibfnamefont {J.}~\bibnamefont
  {Leach}}, \bibinfo {author} {\bibfnamefont {J.}~\bibnamefont {Courtial}},
  \bibinfo {author} {\bibfnamefont {K.}~\bibnamefont {Skeldon}}, \bibinfo
  {author} {\bibfnamefont {S.~M.}\ \bibnamefont {Barnett}}, \bibinfo {author}
  {\bibfnamefont {S.}~\bibnamefont {Franke-Arnold}}, \ and\ \bibinfo {author}
  {\bibfnamefont {M.~J.}\ \bibnamefont {Padgett}},\ }\href@noop {} {\bibfield
  {journal} {\bibinfo  {journal} {Phys. Rev. Lett.}\ }\textbf {\bibinfo
  {volume} {92}},\ \bibinfo {pages} {013601} (\bibinfo {year}
  {2004})}\BibitemShut {NoStop}%
\bibitem [{\citenamefont {{Tamm}}\ and\ \citenamefont
  {{Weiss}}(1990)}]{Tamm1990}%
  \BibitemOpen
  \bibfield  {author} {\bibinfo {author} {\bibfnamefont {C.}~\bibnamefont
  {{Tamm}}}\ and\ \bibinfo {author} {\bibfnamefont {C.}~\bibnamefont
  {{Weiss}}},\ }\href {\doibase 10.1364/JOSAB.7.001034} {\bibfield  {journal}
  {\bibinfo  {journal} {J. Opt. Soc. Amer.~B}\ }\textbf {\bibinfo {volume}
  {7}},\ \bibinfo {pages} {1034} (\bibinfo {year} {1990})}\BibitemShut
  {NoStop}%
\bibitem [{\citenamefont {Beijersbergen}\ \emph {et~al.}(1993)\citenamefont
  {Beijersbergen}, \citenamefont {Allen}, \citenamefont {der Veen},\ and\
  \citenamefont {Woerdman}}]{Beijersbergen1993}%
  \BibitemOpen
  \bibfield  {author} {\bibinfo {author} {\bibfnamefont {M.~W.}\ \bibnamefont
  {Beijersbergen}}, \bibinfo {author} {\bibfnamefont {L.}~\bibnamefont
  {Allen}}, \bibinfo {author} {\bibfnamefont {H.~V.}\ \bibnamefont {der Veen}},
  \ and\ \bibinfo {author} {\bibfnamefont {J.~P.}\ \bibnamefont {Woerdman}},\
  }\href@noop {} {\bibfield  {journal} {\bibinfo  {journal} {Opt. Commun.}\
  }\textbf {\bibinfo {volume} {96}},\ \bibinfo {pages} {123} (\bibinfo {year}
  {1993})}\BibitemShut {NoStop}%
\bibitem [{\citenamefont {Sullivan}(1972)}]{Sullivan1972}%
  \BibitemOpen
  \bibfield  {author} {\bibinfo {author} {\bibfnamefont {D.~L.}\ \bibnamefont
  {Sullivan}},\ }\href@noop {} {\bibfield  {journal} {\bibinfo  {journal}
  {Appl. Opt.}\ }\textbf {\bibinfo {volume} {11}},\ \bibinfo {pages} {2028}
  (\bibinfo {year} {1972})}\BibitemShut {NoStop}%
\bibitem [{\citenamefont {Rubino}\ \emph
  {et~al.}(2017{\natexlab{b}})\citenamefont {Rubino}, \citenamefont {Rozema},
  \citenamefont {Massa}, \citenamefont {Ara{\'u}jo}, \citenamefont {Zych},
  \citenamefont {Brukner},\ and\ \citenamefont {Walther}}]{rubino2017b}%
  \BibitemOpen
  \bibfield  {author} {\bibinfo {author} {\bibfnamefont {G.}~\bibnamefont
  {Rubino}}, \bibinfo {author} {\bibfnamefont {L.~A.}\ \bibnamefont {Rozema}},
  \bibinfo {author} {\bibfnamefont {F.}~\bibnamefont {Massa}}, \bibinfo
  {author} {\bibfnamefont {M.}~\bibnamefont {Ara{\'u}jo}}, \bibinfo {author}
  {\bibfnamefont {M.}~\bibnamefont {Zych}}, \bibinfo {author} {\bibfnamefont
  {{\v{C}}.}~\bibnamefont {Brukner}}, \ and\ \bibinfo {author} {\bibfnamefont
  {P.}~\bibnamefont {Walther}},\ }\href@noop {} {\  (\bibinfo {year}
  {2017}{\natexlab{b}})},\ \Eprint {http://arxiv.org/abs/1712.06884}
  {arXiv:1712.06884 [quant-ph]} \BibitemShut {NoStop}%
\bibitem [{\citenamefont {Fedrizzi}\ \emph {et~al.}(2007)\citenamefont
  {Fedrizzi}, \citenamefont {Herbst}, \citenamefont {Poppe}, \citenamefont
  {Jennewein},\ and\ \citenamefont {Zeilinger}}]{Fedrizzi:07}%
  \BibitemOpen
  \bibfield  {author} {\bibinfo {author} {\bibfnamefont {A.}~\bibnamefont
  {Fedrizzi}}, \bibinfo {author} {\bibfnamefont {T.}~\bibnamefont {Herbst}},
  \bibinfo {author} {\bibfnamefont {A.}~\bibnamefont {Poppe}}, \bibinfo
  {author} {\bibfnamefont {T.}~\bibnamefont {Jennewein}}, \ and\ \bibinfo
  {author} {\bibfnamefont {A.}~\bibnamefont {Zeilinger}},\ }\href {\doibase
  10.1364/OE.15.015377} {\bibfield  {journal} {\bibinfo  {journal} {Opt.
  Express}\ }\textbf {\bibinfo {volume} {15}},\ \bibinfo {pages} {15377}
  (\bibinfo {year} {2007})},\ \Eprint {http://arxiv.org/abs/0706.2877}
  {arXiv:0706.2877 [quant-ph]} \BibitemShut {NoStop}%
\bibitem [{\citenamefont {Erhard}\ \emph {et~al.}(2018)\citenamefont {Erhard},
  \citenamefont {Fickler}, \citenamefont {Krenn},\ and\ \citenamefont
  {Zeilinger}}]{erhard2018twisted}%
  \BibitemOpen
  \bibfield  {author} {\bibinfo {author} {\bibfnamefont {M.}~\bibnamefont
  {Erhard}}, \bibinfo {author} {\bibfnamefont {R.}~\bibnamefont {Fickler}},
  \bibinfo {author} {\bibfnamefont {M.}~\bibnamefont {Krenn}}, \ and\ \bibinfo
  {author} {\bibfnamefont {A.}~\bibnamefont {Zeilinger}},\ }\href@noop {}
  {\bibfield  {journal} {\bibinfo  {journal} {Light: Science \& Applications}\
  }\textbf {\bibinfo {volume} {7}},\ \bibinfo {pages} {17146} (\bibinfo {year}
  {2018})}\BibitemShut {NoStop}%
\bibitem [{\citenamefont {Cerf}\ \emph {et~al.}(2002)\citenamefont {Cerf},
  \citenamefont {Bourennane}, \citenamefont {Karlsson},\ and\ \citenamefont
  {Gisin}}]{cerf2002security}%
  \BibitemOpen
  \bibfield  {author} {\bibinfo {author} {\bibfnamefont {N.~J.}\ \bibnamefont
  {Cerf}}, \bibinfo {author} {\bibfnamefont {M.}~\bibnamefont {Bourennane}},
  \bibinfo {author} {\bibfnamefont {A.}~\bibnamefont {Karlsson}}, \ and\
  \bibinfo {author} {\bibfnamefont {N.}~\bibnamefont {Gisin}},\ }\href@noop {}
  {\bibfield  {journal} {\bibinfo  {journal} {Phys. Rev. Lett.}\ }\textbf
  {\bibinfo {volume} {88}},\ \bibinfo {pages} {127902} (\bibinfo {year}
  {2002})}\BibitemShut {NoStop}%
\end{thebibliography}%

%

\end{document}